\documentclass[a4paper]{llncs}

\usepackage{tabu}
\usepackage{listings}

\newcommand{\sref}[1]{Section~\ref{#1}}
\newcommand{\fref}[1]{Figure~\ref{#1}}

\begin{document}
\title{Measuring the impact of input data on energy consumption of software}
\author{Jeremy Morse}
\institute{University of Bristol}

\maketitle

\begin{abstract}
The amount of energy consumed during the execution of software, and the ability
to predict future consumption, is an important factor in the design of embedded
electronic systems. In this technical report I examine factors
in the execution of software that can affect energy consumption. Taking a
simple embedded software benchmark I measure to what extent input data can
affect energy consumption, and propose a method for reflecting this in software
energy models.
\end{abstract}

\section{Introduction}

Energy consumption in embedded devices is a significant challenge in system
design, with slow improvement in energy storage technology checking the
increasing computational demands on devices. While the research community
continues to study energy-specific software optimisations,
understanding how different portions of software relate to the amount of energy
consumed is of interest to engineers, for example to apply Amdahl's
law to energy consumption.

Existing operational models of processors allow us to model their consumption of
energy in terms of an always-present base cost, and the effect of individual
instruction interpretation as the processor executes a program
\cite{TiwariWolfeInstructionLevelPowerAnalysi:1996}.
With sufficient information about instruction costs, a model can be built to
accurately predict the energy consumption of a particular trace of
instructions. These models, however, only consider an overall or average case
valuation of instruction costs, with no regard for instructions that may
consume different amounts of energy in different circumstances. As we shall
see in this report, the variation in energy consumption caused by
operands to instructions can be significant.

This limitation undermines the accuracy of energy consumption
analysis techniques, used to predict the energy consumption of software.
To provide a safe upper bound on the amount of energy that a particular
sequence of instructions will consume,
one must assume that each instruction consumes the
maximum amount of energy that it possibly could.
This provides an overestimate of the maximum amount of energy a piece of
software may consume. Conversely, considering the average cost of each
instruction may give a more realistic prediction of the software's normal
energy consumption, but gives no guarantee that consumption will not
exceed that amount in other circumstances. This is a difficulty shared with
the \textit{worst case execution time} (WCET) problem \cite{wcet-intro-survey},
where the longest possible path through a program must be identified, although
our focus is the consumption of energy rather than the consumption of time.

To provide a tighter bound on the worst case energy consumption of software,
I propose the use of simple static analyses to identify instruction which can
exhibit worst-case energy consumption, those that cannot, and to compute costs
for each instruction appropriately. This report is organised as follows:
\sref{sec:background} covers the background to energy analysis of software.
\sref{sec:experiments} studies the energy impact of input data on a benchmark
running on a XMOS XS1-L processor. \sref{sec:models} examines existing software
static analysis techniques and how they can be applied to energy consumption.
\sref{sec:futurework} draws conclusions and outlines my future work.

\section{Background to energy modelling}
\label{sec:background}

At the lowest level, the consumption of energy by a processor is caused by the
charging of internal circuitry performing calculations or moving data during
the execution of an instruction. Tiwari et al.
\cite{TiwariWolfeInstructionLevelPowerAnalysi:1996} characterised
the costs involved by classifying three energy consuming operations: the base
cost of executing a particular instruction, the cost of switching the processor
from executing one kind of instruction to another, and miscellaneous other
costs. Tiwari summaries computes these for a particular trace of instructions
executed, as

\begin{equation}
    E_p = \sum_{i} \left( B_i \times N_i \right) +
    \sum_{i,j} \left( O_{i,j} \times N_{i,j} \right) +
    \sum_{k} E_k
    \label{eq:tiwari}
\end{equation}

Where $N_i$ is the number of times instruction $i$ is executed, $B_i$ the
energy cost of executing that instruction, $N_{i,j}$ the number of transitions
between instruction $i$ to $j$ and $O_{i,j}$ the cost of the same transition.
$E_k$ represents miscellaneous other costs, while $E_p$ is the total cost for
the whole trace. Within this formulation there are two distinct parts: the
modelling of energy costs for actions (such as executing an instruction), and
the count of how many times actions are performed. The former constitutes the
energy model of the processor, allowing the calculation of an energy cost for
any given trace of instructions, while the latter is derived from a particular
execution of a program, the mechanisms for which are not considered here.

Such an energy model \cite{steve-energy-model} has been produced for the
XMOS XS1-L processor. The XCore architecture \cite{XMOS2009a} centres around
a RISC execution core running multiple hardware threads, connected to other
cores via high-speed serial links. The XCore was designed to be predictable
and deterministic, avoiding superscalar execution, branch prediction and memory
caches. It
schedules threads to run in a fixed (round-robin) order. It's predictability
makes the XCore particularly suitable for performing
and evaluating modelling, without risk of interference.

The XS1-L energy model \cite{steve-energy-model} is comprised of attributes for a core set of
instructions for which the measurements prescribed by Tiwari et al. have been
made. These core instructions have then been characterised and the costs
fitted to other instructions for which measurements cannot be made.
Accommodation is also made for inter-instruction costs ($O_{i,j}$ in the
formulation above) and the effects of concurrently executing instructions
from different threads.

This energy model has been leveraged in \cite{lopstr-paper} to allow energy
consumption analysis of instruction traces, providing a metric for use with software cost analysis as
provided by the Ciao framework \cite{ciao-overview}. The result is a formula
for calculating the average energy consumption of the program, given a
particular size of input. While there is much research on the topic of generic
resource analysis, very little work has been done in the field of energy
analysis, specifically with regard to analysis for \textit{worst
case} energy consumption (WCEC), the discovery of safe upper bounds on the programs
energy consumption. In particular, I am not aware of any work on the energy
consumption of programs with specific regard to the data that they operate
upon. This will be studied further in \sref{sec:experiments}.

A substantial amount of work has gone into the study of the WCET problem
\cite{wcet-intro-survey}, which does vary with the input data provided to
programs. The distinguishing feature is one of \textit{size}: inputs provided
to a program that affect the amount of code run (such as number of loop
iterations or instructions executed) affect the runtime of the program.
Discovering the ``largest'' input corresponds to finding the longest that the
program can run. In contrast, WCEC considers the variation in input data that
does not affect the length of program paths, but instead affects the amount
of energy consumed by instructions along those paths.

Within the WCET community itself, numerous techniques have been used to
identify the longest program path. A full treatment is given in
\cite{wcet-intro-survey}, however notable techniques include
\textit{implicit path enumeration} where the longest paths from a branch are
identified locally and composed to build a worst-case path, without any
explicit path exploration.
Abstract interpretation \cite{absint} can be used to statically analyse
potential paths through the program and reason about how different code paths
compose.

\section{Impact of data on energy consumption}
\label{sec:experiments}

To correctly model the effect of input data on energy consumption, we must
better understand the effects of variations in input data.
At the lowest level, the \textit{hamming distance} between two
values is the the number of bits that must charge or discharge across the
processor pipeline. We would expect consecutive operations on values with
large hamming distances to result in higher energy consumption than when the
values have small hamming distances. The actual impact such distances can have
on a particular processors energy consumption can only be measured by
experimentation, however.

The work of Tiwari et al. has already established that some instructions cost
more than others during execution. This could be because different instructions
have differing base costs---it could also be because different amounts of
circuitry are switched by each instruction,
in which case energy consumption will
scale with hamming distance by different constants for different instructions.
The energy of input data, compared to base costs, are unknown.

To resolve this matter, I took a simple benchmark for the XMOS XS1-L
processor that computed a finite impulse response (FIR)over a pre-determined set
of input data, and altered it in a number of ways. In all cases I measured the
average energy consumption of the processor during execution of the benchmark.
Firstly, I fed several different patterns of input data into the algorithm with
different hamming distances. Secondly, I altered the core operation of the FIR
benchmark to measure how different instructions scale energy consumption with
input data.

\subsection{Finite impulse response benchmark}

The finite impulse response (FIR) is a basic DSP algorithm to filter an input
signal for certain frequencies, according to it's configuration. The core of
the algorithm is a window of input samples, each sample of which is multiplied
by a coefficient according to it's position in the window. The results of all
multiplications are summed to produce the output signal sample. For each output
sample calculated, the window of input samples is shifted by one.

This core part is shown in Figure~\ref{fig:codesample} written in XC.
The \texttt{xn}
argument contains the newest input sample to be processed, \texttt{state} is
the window of input samples currently being processed, \texttt{coeffs} the
values by which samples are multiplied, \texttt{ELEMENTS} the length of the
input window and \texttt{ynh} \texttt{ynl} are the high and low parts of a
64 bit integer. XC supports multiple values being returned from functions,
which are enclosed in curly braces in the function signature, return statement,
and call site. The main multiply-and-accumulate function of the algorithm is
performed by the \texttt{macs} intrinsic.

\begin{figure}
\begin{lstlisting}
{int,int,int} fir(int xn, const int coeffs[], int state[],
                  int ELEMENTS, int ynh, unsigned int ynl)
{
    int o = state[ELEMENTS-1];
    for(int j=ELEMENTS-1; j!=0; j--) {
        state[j] = state[j-1];
        {ynh, ynl} = macs(coeffs[j], state[j], ynh, ynl);
    }
    state[0] = xn;
    {ynh, ynl} = macs(coeffs[0], xn, ynh, ynl);

    return {o, ynh, ynl};
}
\end{lstlisting}
\label{fig:codesample}
\end{figure}

The XCore architecture features multiple hardware threads.
The processor contains a single execution pipeline, which executes instructions
from each active hardware thread in a round robin schedule. To make full use of
the available resources (see \cite{steve-energy-model} for a full explanation),
the FIR benchmark used here splits it's calculation
into seven stages of equal length, which are then chained together across
seven concurrent threads. The Tiwari energy model explained in 
\sref{sec:background} still applies to concurrent execution on the XCore,
however the transitions between instructions are now also transitions between
threads.

\subsection{Input data and algorithm changes}

Within the FIR benchmark, I vary the values of input samples fed into the
sampling window and used as the multiplication coefficients. The different sets of
values have different hamming distances, controlled  by keeping a fixed number of
leading zero bits in each value. The first set is of random 8 bit numbers, with the preceding
24 bits in each integer clamped to zero. The same approach is used to produce
sets of random 16, 24 and 32 bit numbers. Two other special input sets are
used: a set of all-zero values, and a set of samples from a sine wave with a
period that repeats every 24 samples. Each input set should provide insight
into how different hamming distances affects energy consumption, with the sine
wave signal providing a reference for normal operation of the benchmark.

To measure how different instructions scale their energy consumption with input
samples, I also alter the main operation of the FIR benchmark to use different
instructions. This alteration occurs at the assembly level, to avoid any
unwanted changes introduced by the compiler. By default as shown in 
\fref{fig:codesample} the FIR benchmark multiplies an input sample with a
coefficient and accumulates it into a sum variable. This translates to a
single instruction, \texttt{maccs}. For these measurements, I replace the
\texttt{maccs} instruction with 
\texttt{nop}, \texttt{add}, \texttt{sub}, \texttt{xor} and \texttt{lmul},
representing some common processor operations that have different instruction
costs in the energy model \cite{steve-energy-model}. According to that model,
we would expect \texttt{nop} to not
change energy consumption with input data at all; add, sub and xor to
scale to a lesser extent than the multiply-and-accumulate instruction, and
lmul to use an equivalent or possibly more energy than maccs.

In addition to replacing the base operation of the algorithm, I also varied
two other factors. First, I repeated the main operation several times,
duplicating the \texttt{maccs} instruction (or otherwise) from one to seven
times. I also repeated all my tests with the core instruction operands rewritten
to operate on a fixed, low hamming distance piece of data, in this case the
loop iterator.\footnote{which ranges from 1 to 18}
These tests will allow comparison between instructions executing
on high hamming distance constantly changing input and low range operands.

\subsection{Test setup}
\label{sec:testsetup}

All tests were run on a SliceKit Analogue development board \cite{slicekit},
with connected XTag programmer board.
Energy consumption was measured in the usual way, with current-sense samples
directly recorded by the XTag programmer. The average current draw was
determined by sampling a 40us period during the middle of the benchmark
execution and taking the average amount of current over that period. All
readings are reported in milliWatts. Each result is averaged over 3 individual
test-runs, to reduce the effect of any noise introduced during current
readings.

The SliceKit itself was configured to run at 400Mhz and with default 3.3 Volt
supply. This report does not consider DVFS, and so these parameters are not
modified. With the XCore idling (one thread blocking on a never-triggered
event) the processor consumes 200mW in this configuration. This should be
considered to be the baseline amount of power consumption: the amount over
this rate represents the contribution of software to the processors energy
consumption.

\subsection{Results}

The results of the first experiment are presented in \fref{tab:results}. Each
row represents the average power consumption (in milliwats) of the FIR
benchmark using the instruction given in the leftmost column. The other columns
represents the energy readings for each input pattern.

\begin{figure}
\begin{tabu}to \textwidth{|c||X[c]|X[c]|X[c]|X[c]|X[c]|X[c]|}
\hline
Instruction & \multicolumn{6}{c|}{Input patterns} \\
\hline
 & zeros & rand8 & rand16 & rand24 & rand32 & signal \\
\hline
maccs & 218.79 & 223.24 & 228.93 & 233.65 & 238.28 & 234.65 \\
\hline
lmul & 219.95 & 224.29 & 229.49 & 233.95 & 239.09 & 234.69 \\
\hline
sub & 220.07 & 226.28 & 228.72 & 231.12 & 233.50 & 230.81 \\
\hline
add & 220.30 & 223.15 & 226.96 & 230.00 & 233.33 & 230.81 \\
\hline
xor & 219.63 & 223.57 & 228.94 & 232.91 & 237.49 & 234.17 \\
\hline
nops & 218.52 & 219.84 & 220.83 & 221.88 & 223.68 & 222.41 \\
\hline
\end{tabu}
\caption{Milliwat consumption of Analogue SliceKit XCore running FIR benchmark with the given instruction and input pattern}
\label{tab:results}
\end{figure}

\fref{tab:results2} and \fref{tab:results3} present the results of the
additional tests I ran, increasing the number of times the main instruction
of the algorithm are executed per iteration. \fref{tab:results2} shows that as
we increase the number of times the core operation of the algorithm executes,
the energy consumption of the benchmark increases. This is in line with
expectations, as the more frequently an expensive instruction is executed,
the greater the amount of energy consumed.

\fref{tab:results3} compares a similar scaling of the number of times the core
instruction is executed, but comparing the energy consumption when the
instruction operates on a fixed piece of data, and when it operates upon the
input samples. This is signified by ``not in dpath'' and ``in dpath'' in the
instruction description, respectively.
We can clearly see that energy consumption is higher
when the instructions operate on the input data rather than data of limited
range.

\begin{figure}
\begin{tabu}to \textwidth{|c||X[c]|X[c]|X[c]|X[c]|X[c]|X[c]|X[c]|}
\hline
Instruction & \multicolumn{7}{c|}{Insn repetition from 1 to 7} \\
\hline
 & 1 & 2 & 3 & 4 & 5 & 6 & 7 \\
\hline
maccs & 239.38 & 246.75 & 252.11 & 252.39 & 253.62 & 254.54 & 255.10\\
\hline
lmul & 238.88 & 246.41 & 251.40 & 252.15 & 252.76 & 254.00 & 254.46\\
\hline
add & 233.31 & 237.92 & 241.86 & 245.06 & 247.87 & 249.73 & 249.99\\
\hline
sub & 233.48 & 237.95 & 242.18 & 245.63 & 247.80 & 250.24 & 250.53\\
\hline
xor & 237.39 & 243.97 & 248.72 & 248.94 & 249.38 & 250.21 & 250.73\\
\hline
nops & 223.98 & 224.08 & 223.99 & 224.01 & 224.04 & 224.09 & 223.90\\
\hline
\end{tabu}
\caption{Milliwat consumption of FIR benchmark with varying repetition of core instruction, when fed the rand32 input pattern}
\label{tab:results2}
\end{figure}

\begin{figure}
\begin{tabu}to \textwidth{|c||X[c]|X[c]|X[c]|X[c]|X[c]|X[c]|X[c]|}
\hline
Instruction & \multicolumn{7}{c|}{Insn repetition from 1 to 7} \\
\hline
 & 1 & 2 & 3 & 4 & 5 & 6 & 7 \\
\hline
maccs in dpath & 239.38 & 246.75 & 252.11 & 252.39 & 253.62 & 254.54 & 255.10 \\
\hline
maccs not in dpath & 232.67 & 235.09 & 237.76 & 237.65 & 237.01 & 236.87 & 236.28 \\
\hline
lmul in dpath & 238.88 & 246.41 & 251.40 & 252.15 & 252.76 & 254.00 & 254.46 \\
\hline
lmul not in dpath & 230.47 & 231.88 & 234.00 & 233.82 & 232.58 & 232.45 & 231.68 \\
\hline
\end{tabu}
\caption{Milliwat consumption of FIR benchmark with varying repetition of core instruction, when fed the rand32 input pattern, varying whether the instruction operates on fixed or input data}
\label{tab:results3}
\end{figure}

\subsection{Discussion}

As expected, an increasing hamming distance between values (corresponding to
smaller zero-bit prefixes for the random samples) results in higher energy
consumption in all operations. The greatest increase is for the \texttt{lmul}
instructions, rising from 219mW when operating on all-zero inputs to 239mW
when fed random 32 bit values. This represents only 10\% of the overall energy
consumption of the system, but increases the software contribution to energy
consumption by 100\%.\footnote{Taking 200mW as the base cost, as discussed in
\sref{sec:testsetup}}

The \texttt{nop} instruction exhibits the smallest increase in energy
consumption as input patterns change. This is no surprise, as the instruction
does not actually manipulate any data. Examining the assembly of the core loop
in the FIR benchmark, shown in \fref{fig:assemblylisting}, where r4 references the
\texttt{state} array and r2 the \texttt{coeffs} array, we see that the only
instructions accessing the input data are load and store instructions. It is
logical to assume that the 6mW difference between the all-zeros and rand32
input patterns when using \texttt{nop} as the benchmark operation is due to
those loads and stores.

\begin{figure}
\begin{lstlisting}
.LBB0_1:
	sub r9, r8, 1
	ldw r10, r4[r9]
	stw r10, r4[r8]
	ldw r8, r2[r8]
	nop
	mov r8, r9
	bt r9, .LBB0_1
\end{lstlisting}
\caption{Core loop of FIR benchmark when using \texttt{nop} instruction instead of \texttt{maccs}}
\label{fig:assemblylisting}
\end{figure}

We see that the ``signal'' input to the FIR benchmark, representing a typical
input for the algorithm in a real application, consistently results in a lower
rate of energy consumption than the random 32-bit samples. This too meets
with expectations: the sine wave follows an oscillating pattern that slowly
moves from high integer values to low (crossing zero into the negative range)
over the period of the wave. This keeps some of the higher order bits of each
sample the same for several samples, reducing the hamming distance.

The base cost of each instruction when no data is operated upon (i.e., all
inputs are zero) are roughly equal. There is a small difference between certain
instruction (\texttt{maccs} and \texttt{nop} being 1.5mW lower than \texttt{add}
and \texttt{sub} for example), however these differences are less than 1\% of
the overall energy cost of the processor. This seems to confirm that there is
little or no base cost to each instruction itself, and the increased
consumption scales up with increased hamming distance of data. The scale-up
appears linear with instructions such as \texttt{maccs}, \texttt{lmul},
\texttt{add} and \texttt{xor}, but not for \texttt{sub}.
This is because the second operand to subtracts in twos-compliment arithmetic
are inverted, increasing the hamming distance between operands.

Considering the results in \fref{tab:results2} we see
that as more instructions operating on data are added to the core loop of the
algorithm, the energy consumption increases, in line with expectations. The
increase is not linear, and flattens out when the instruction reaches 7
repetitions.\footnote{Repetitions past 7 are not presented here} This is presumably because the data operation instruction occurs
as frequently as the other instructions in the loop, see
\fref{fig:assemblylisting}.

\fref{tab:results3} also shows that a substantial portion of the instruction
energy cost depends on the data that it operates on, as performing an operation
on the low-range data (the loop iterator, ranging from 1 to 18) consumes less
than operating on the random 32 bit input data. There is a discrepancy with the
previous results however, as we would expect the energy consumption when each
instruction is repeated once to match the ``rand8'' consumption reading from
\fref{tab:results}. Instead, more energy is consumed in this setup. This
amounts to approximately 2\% of the total reading. One potential explanation
is that the load and store operations around the operation instruction, which
are still loading and storing the random 32-bit data, may contribute the
additional energy cost. Regardless, the difference between operating on data
in the datapath and not, is shown to be significant.

\section{Worst and average case energy models}
\label{sec:models}

These results illustrating how energy consumption scales with input data
provide a basis for refining the processor energy model. The main observation
is that almost all the increase in energy consumption as input data hamming
distance increases, is controlled by which instruction is used for the FIR
calculation. If \texttt{maccs} is used, consumption scales up significantly,
which if \texttt{nop} is used, it does not. This can be generalised into the
observation that we only need to consider the worst case energy consumption
for instructions that may operate on data with the greatest hamming distance.

We can then classify instructions into two broad classes: those that operate on
data with the greatest hamming distance, and those that do not. The naive
approach would be to explore every path through the program with every possible
input, and compute the hamming distance for every instruction. This would
immediately result in \textit{state space explosion} \cite{modelcheckingbook},
making
analysis of any non-trivial programs infeasible. The corollary is that we
cannot compute the most energy-consuming data that a particular instruction in
a program may operate upon, as that would require exploring the program to find
the input that leads to that situation \cite{wcet-intro-survey}.

To feasibly analyse a program, we must make approximations of it's inputs and
reason about whether they may lead to worst-case operands for an instruction
\cite{wcet-intro-survey}.
Rather than explicitly explore all the inputs to a program, we may instead
classify any input data at all\footnote{i.e., values read from a peripheral,
communication stream, or other external source} as potentially having a
worst-case value, for any instruction that operates upon it. This reduces
accuracy, as some operations of the analysed program may reduce the hamming
distance of an input, but means that we can use static analysis techniques to
identify instructions that operate on input data.

Specifically, we may use existing data flow analyses \cite{data-flow-analysis}
such as abstract interpretation \cite{absint}
or taint analysis \cite{taintanalysis} to identify
instructions that read in input data, track where that data flows through the
rest of the program, and which instructions operate upon the data. These
instructions may consume the worst-case amount of energy. At the same time,
however, the instructions that do not operate on input data must also be
classified. Not operating on data, these instructions all maintain state
internal to the program, for example counters and loop iterators, ringbuffer
pointers, and so forth. These pieces of data may posses a significant range of
values, however as they are entirely internal to the program we can statically
determine their values, for example through an interval analysis that
identifies upper and lower bounds on data values.

This instruction classification would allow us to identify the approximate
inputs to each instruction in the program, and as a result we could select an
appropriate energy cost valuation for each---assuming such an energy model is
available. While this analysis would not be completely
precise, it avoids having to use the worst case energy cost for every
instruction, and would thus lead to a tighter worst-case bound on energy
consumption.

\section{Conclusions and future work}
\label{sec:futurework}

This technical report has studied the relation between input data to a software
algorithm and the energy consumption of that algorithm, showing that energy
consumption of software grows as the hamming distance of inputs
grows. In certain cases the contribution of data to the dynamic energy
consumption of the program can be 100\%.
I make observations about how input data affects different sets of
instructions in a program, and propose analyses to classify instructions into
sets that may consume the worst-case amount of data, and those that consume
less.

In future work, I will fully implement the proposed analyses and evaluate their
impact on making predictions about the energy consumption of software. To date
I have used a taint analysis within the KLEE \cite{klee} symbolic execution
tool to identify input data manipulating instructions, with promising results
on some simple benchmarks. The interval analysis of internal state manipulating
instructions is yet to be implemented.

\bibliographystyle{plain}
\bibliography{wcenergy}

\end{document}